\documentclass{PoS}

\usepackage{epsfig}
\usepackage{subfigure}

\title{Using Faraday Rotation Gradients to probe Magnetic Tower Models}

\ShortTitle{Faraday Rotation Gradients in AGN}

\author{\speaker{Mehreen Mahmud}\\

        University College Cork, Ireland\\

        E-mail: \email{mahmud@physics.ucc.ie}}

\author{Denise C. Gabuzda\\

        University College Cork, Ireland\\

        E-mail: \email{gabuzda@physics.ucc.ie}}

\abstract{Parsec-scale multi-wavelength VLBA polarization observations can 
be used to study the magnetic-field structures of Active Galactic Nuclei (AGN) 
based on Faraday Rotation (FR) gradients. A number of transverse FR gradients 
have been found, and interpreted as corresponding to helical magnetic fields 
wrapped around the jets; the gradients reflect the systematic change in the 
line-of-sight component of a toroidal or helical magnetic field across the jet 
\cite{Asa02,Gab04,Zav05}. Our observations of a sample of BL Lac objects at six 
wavelengths near 2, 4 and 6 cm have also revealed a previously undetected 
phenomena: these transverse gradients sometimes change their direction with 
distance from the core. We have observed this behaviour in at least five sources,
which display gradients in their VLBI core region opposite to those in the jet.
We suggest that this may be linked to magnetic tower models. In magnetic tower 
models, the field lines go outward with the jet and return and close in the 
accretion disk (or vice versa); differential rotation of the accretion disk 
winds up the inner and outer field lines into two helices (the inner helix 
``nested'' in the outer helix). The total observed FR gradient is a sum of the 
effect of these two helical fields. It may be that gradients detected relatively
far from the core correspond to the outer helix, while gradients detected in the
core region correspond to dominance of the inner helix. This provides tentative 
evidence for the unification of helical magnetic fields and magnetic tower models, which could provide crucial new information for understanding AGN jets.
Further VLBI studies with resolution sufficient to reliably detect these 
gradients in the cm-wavelength core and inner jet will be important for further
investigations of this phenomena.}

\FullConference{The 9th European VLBI Network Symposium on The role of VLBI in 
the Golden Age for Radio Astronomy and EVN Users Meeting\\

		 September 23-26, 2008\\

		 Bologna, Italy}

\begin{document}

\section{Introduction}

The radio emission of Active Galactic Nuclei (AGN) is synchrotron radiation 
generated in the relativistic jets that emerge from the nucleus of the galaxy,
presumably along the rotational axis of a central supermassive black hole.
Synchrotron radiation can be highly linearly polarized, up to $\simeq 75\%$ in
the case of a uniform magnetic ({\bf B}) field \cite{Pach70}. Linear 
polarization observations are essential, as they give information about the 
orientation and degree of order of the {\bf B} field, as well as the 
distribution of thermal electrons and the {\bf B}-field geometry in the 
vicinity of the AGN. VLBI polarization observations of BL~Lac objects have 
shown a tendency for the polarization {\bf E} vectors in the parsec-scale jets 
to be aligned with the local jet direction, so that the predominant {\bf B} 
fields are transverse to the jet \cite{Gab00}. It seems likely that many of 
these transverse {\bf B} fields represent the ordered toroidal component of
the intrinsic B fields of the jets, as discussed by \cite{Gab08}, see also 
references therein.
Faraday Rotation studies are crucial for determining the intrinsic {\bf B} field 
geometries associated with the jets. Faraday Rotation of the plane of linear  
polarization occurs during the passage of an electromagnetic wave through a 
region with free electrons and a magnetic field with a non-zero component along 
the line-of-sight. The amount of rotation is proportional to the integral of
the density of free electrons $n_{e}$ multiplied by the line-of-sight {\bf B} 
field, the square of the observing wavelength $\lambda^{2}$, and various 
physical constants; the coefficient of $\lambda^{2}$  is called the Rotation 
Measure (RM):
\begin{eqnarray}
           \Delta\chi\propto\lambda^{2}\int n_{e} B\cdot dl\equiv RM\lambda^{2}
\end{eqnarray}
The intrinsic polarization angle can be obtained from the following equation:
\begin{eqnarray}
           \chi_{obs} =  \chi_0 + RM \lambda^{2}
\end{eqnarray}
where $\chi_{obs}$ is the observed polarization angle, $\chi_0$ is the 
intrinsic polarization angle observed if no rotation occurred, and $\lambda$ 
is the observing wavelength \cite{Burn66}. Simultaneous multifrequency 
observations thus allow the determination of the RM, as well as identifying 
the intrinsic polarization angles. 
Systematic gradients in the RM have been observed across the parsec-scale jets 
of several AGN, interpreted as reflecting the systematic change in the 
line-of-sight component of a toroidal or helical jet B field across the jet 
\cite{Asa02,Blan93,Gab04,Gab08}. 
\section{Observations}
Very Long Baseline Array (VLBA) polarization observations of the sources 
considered here were carried out as part of our study of a sample of 34 BL Lac objects at six frequencies: 4.6, 5.1, 7.9, 8.9, 12.9 and 15.4 GHz. The BL Lac sample was observed over 5 epochs. Presented here are results for 0138-097, 0256+075, 0716+714, 1334-127, 1749+701 and 2155-152.
Standard tasks in the NRAO AIPS package were used for the amplitude calibration and preliminary phase calibration.
\begin{figure*}
\centering
\begin{tabular}{c}
\epsfig{file=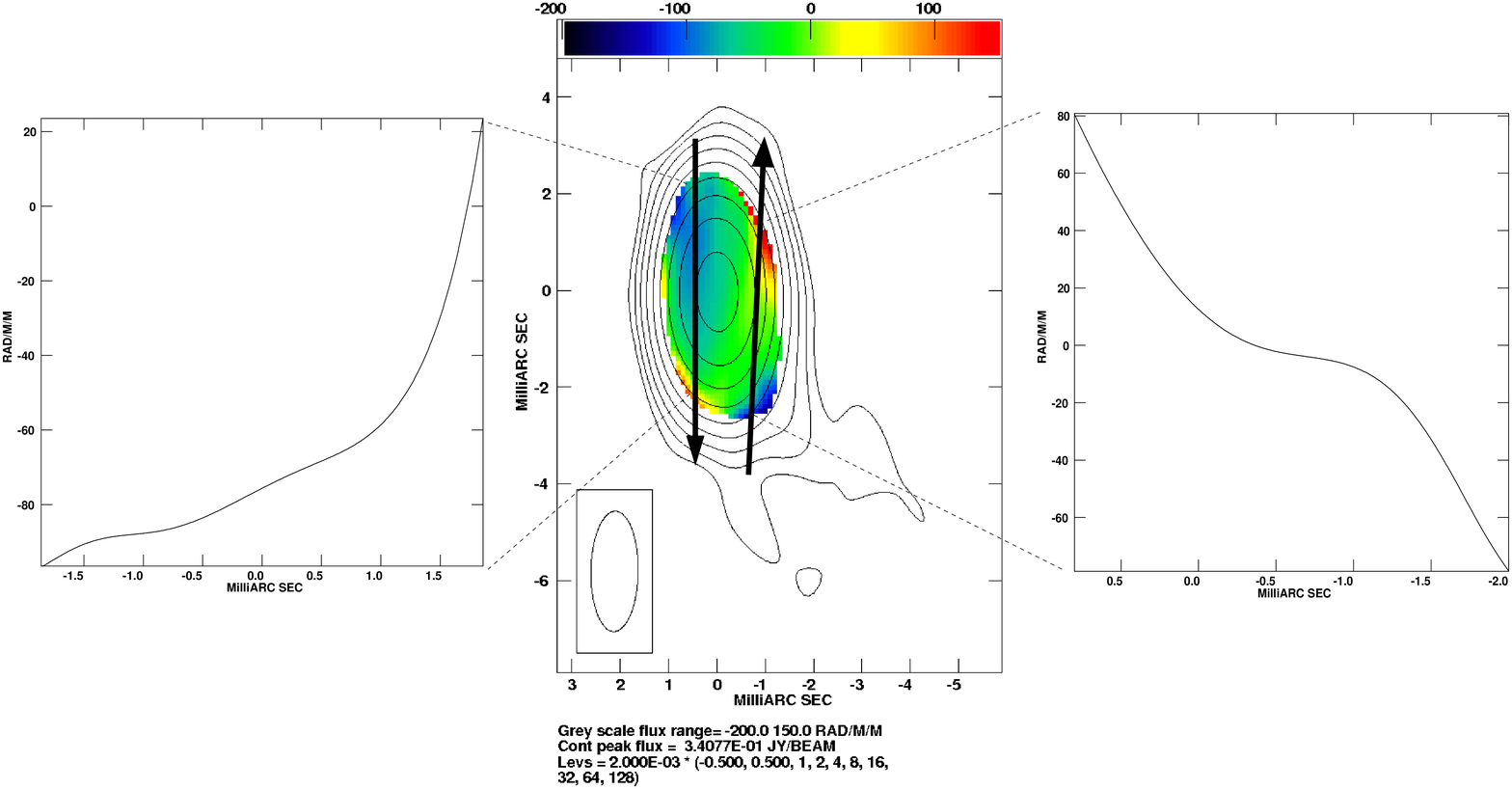,width=0.7\linewidth,clip=} \\
\epsfig{file=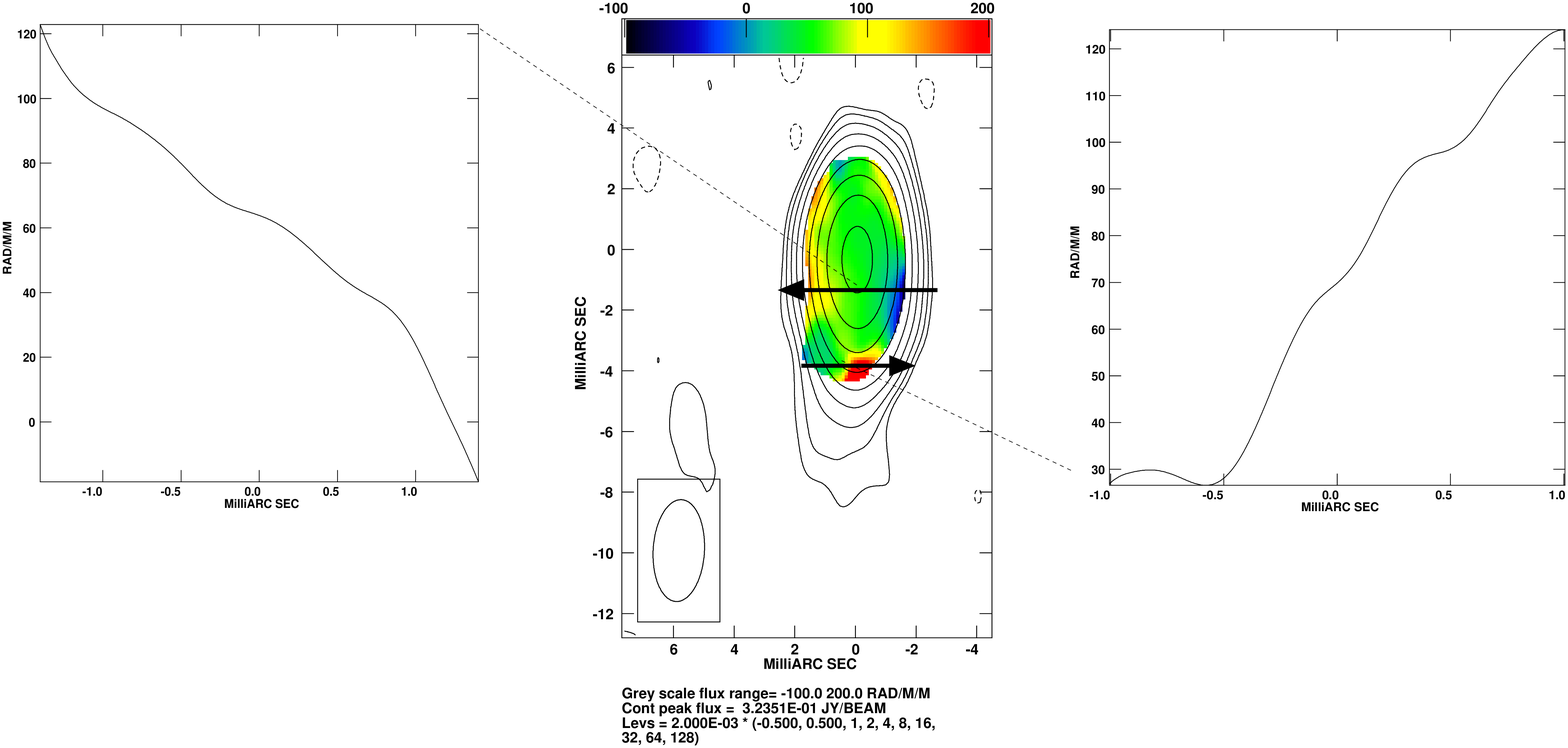,width=0.7\linewidth,clip=} \\
\epsfig{file=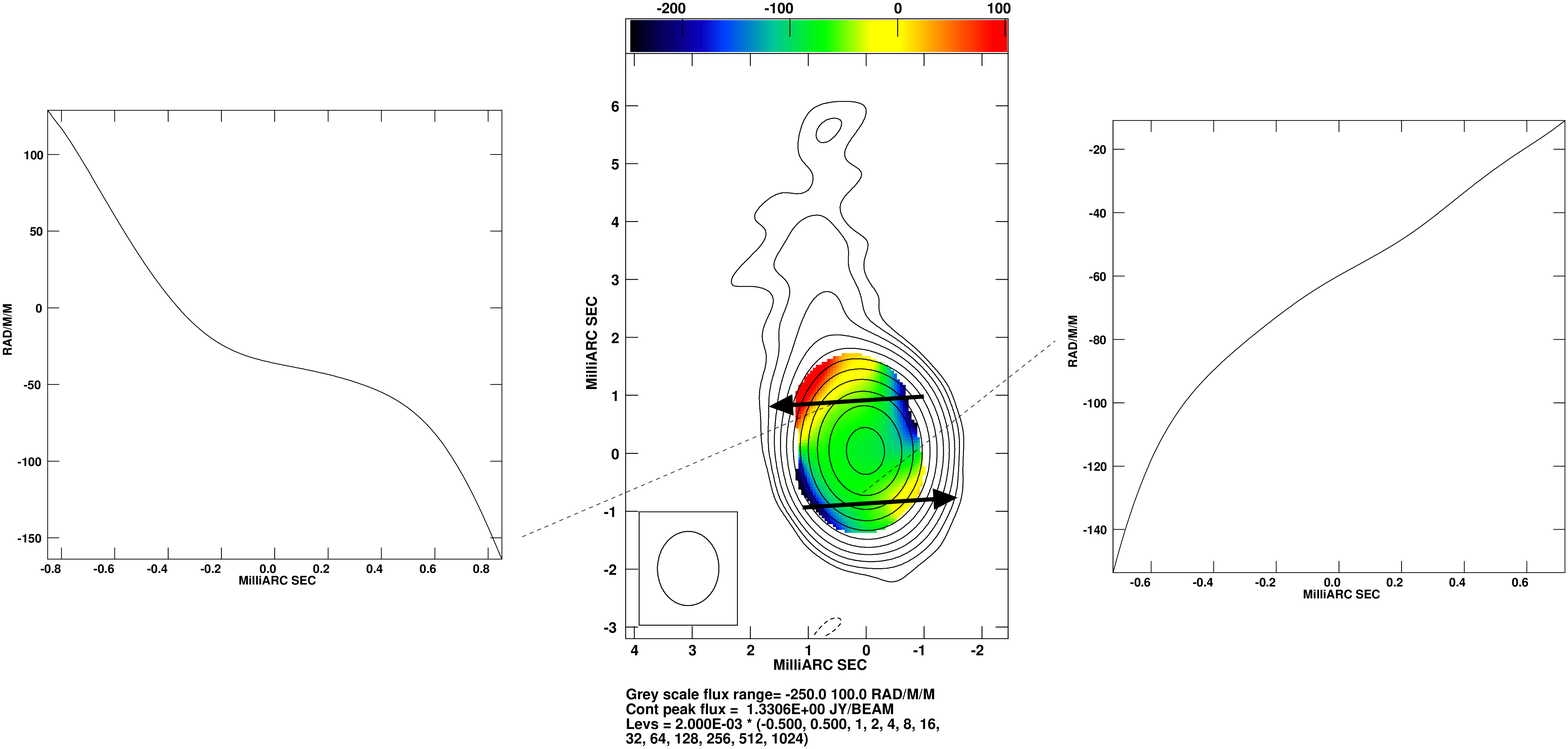,width=0.7\linewidth,clip=} \\
\epsfig{file=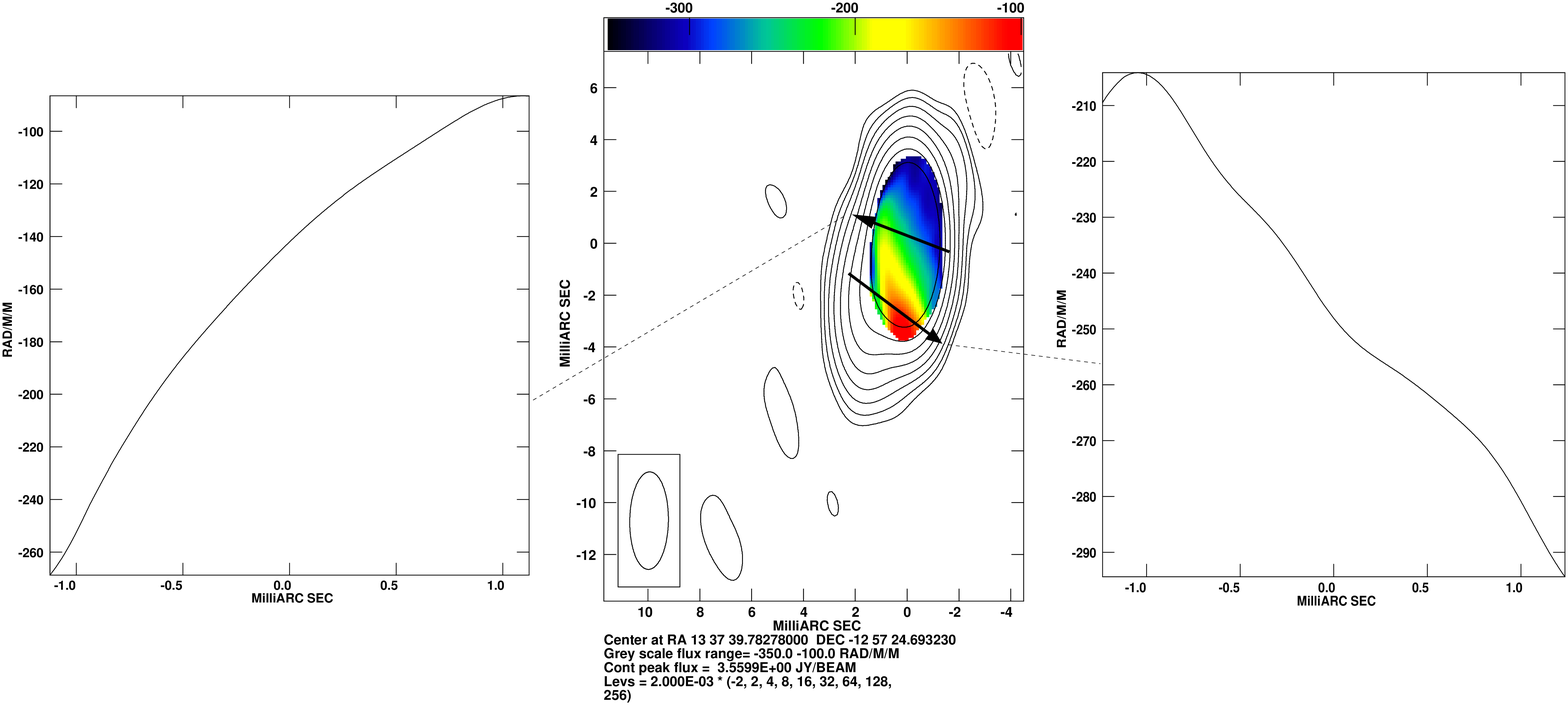,width=0.7\linewidth,clip=} \\
\end{tabular}
\end{figure*}
\begin{figure}
\begin{minipage}[t]{7.0cm}
\begin{center}
\includegraphics[width=5.0 cm,clip]{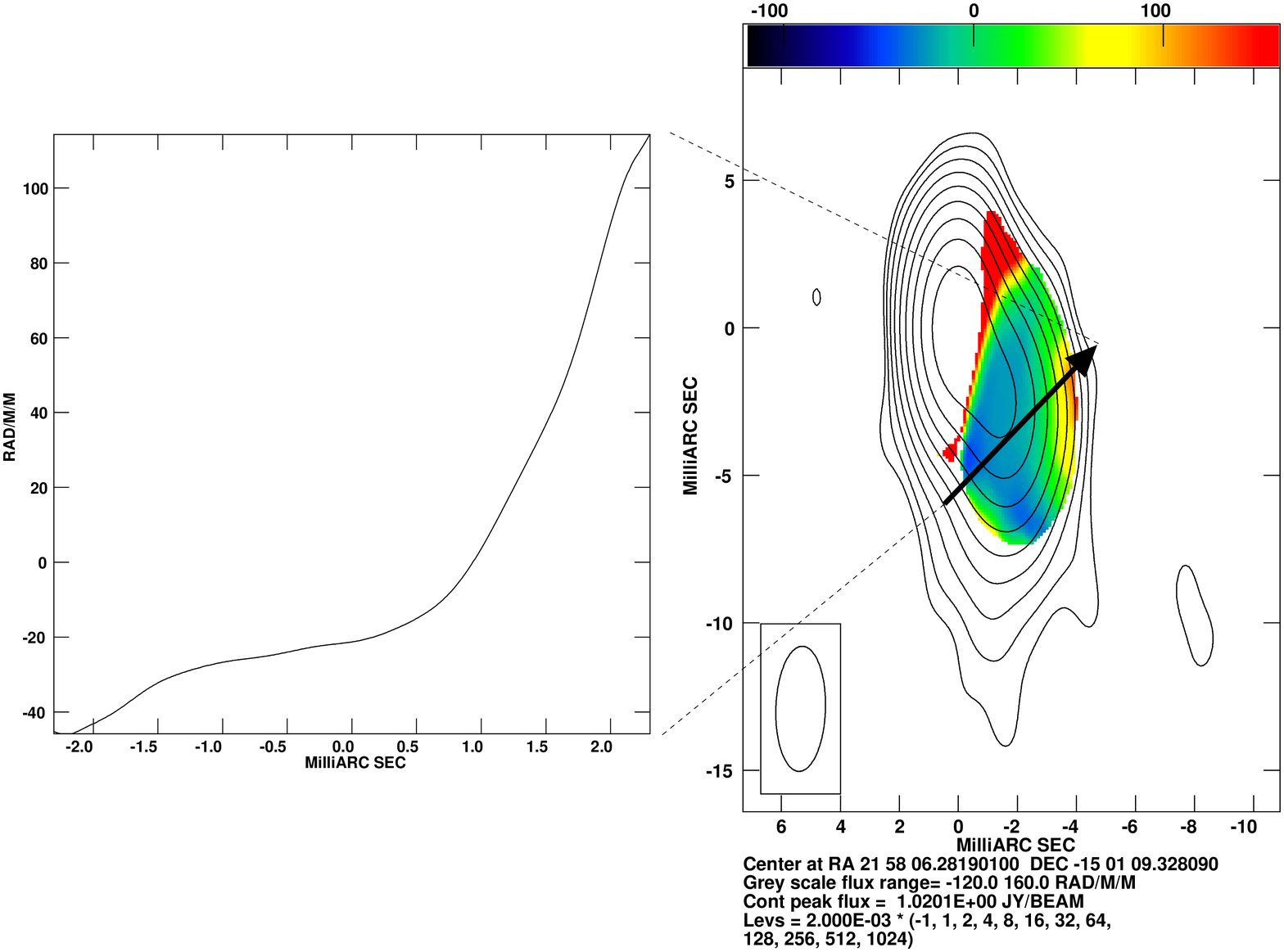}
\end{center}
\vspace{0.5cm}
\end{minipage}
\begin{minipage}[t]{7.0cm}
\begin{center}
\includegraphics[width=5.0 cm,clip]{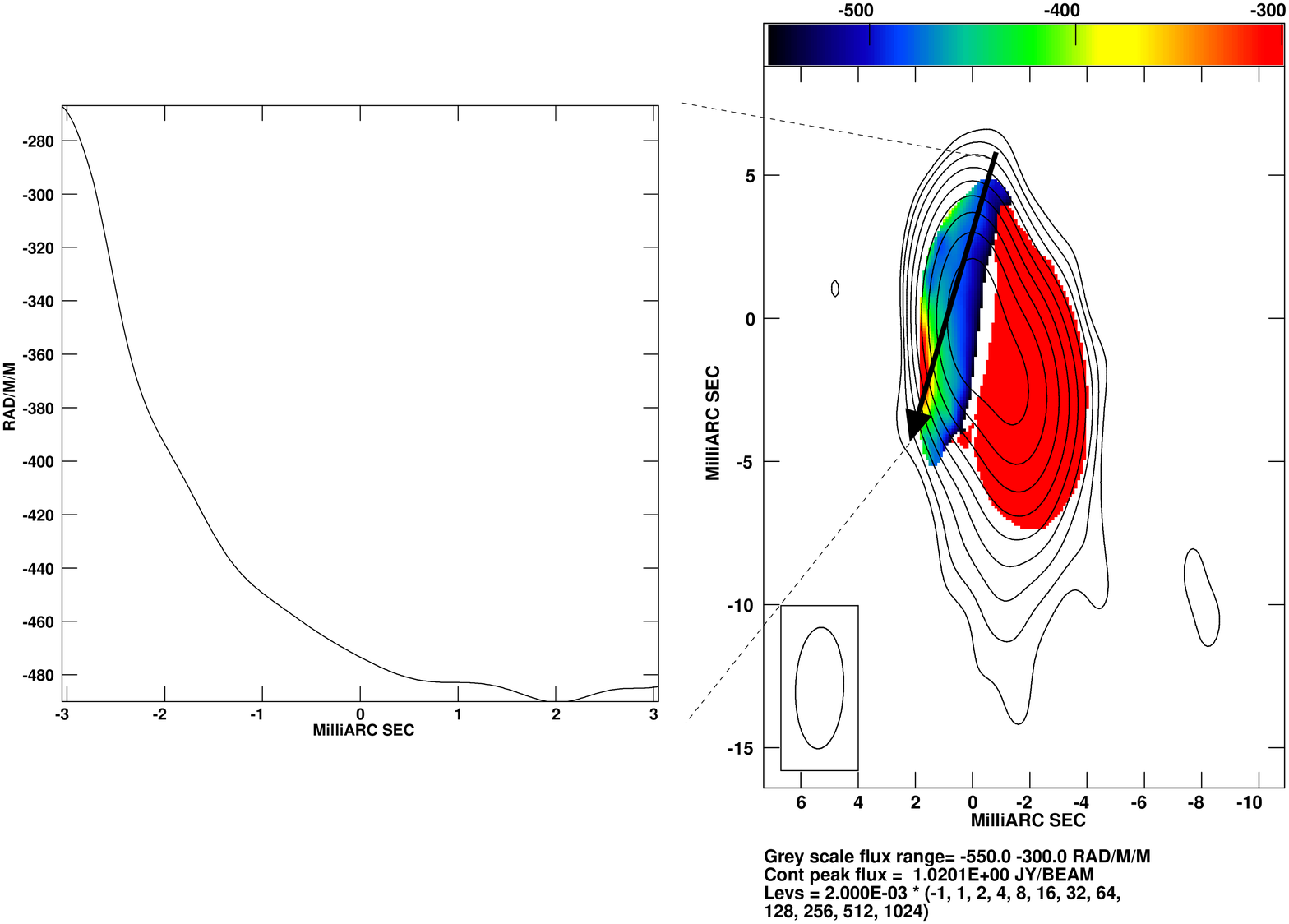}
\end{center}
\end{minipage}
\caption{\label{fig:core_rev2}RM maps of 0138-097, 0256+075, 0716+714, 1334-127 (top to bottom, previous page) and 2155-152 (above left (jet), above right (core)) showing opposing transverse RM gradients in the core and jet. The arrows on the RM maps mark the directions of the gradients. The accompanying panels show slices of the RM distributions across the jets and cores.}
\end{figure}

 \begin{figure}
 \begin{minipage}[t]{0.47\linewidth}
 \begin{center}
 \includegraphics[width=1\linewidth,clip]{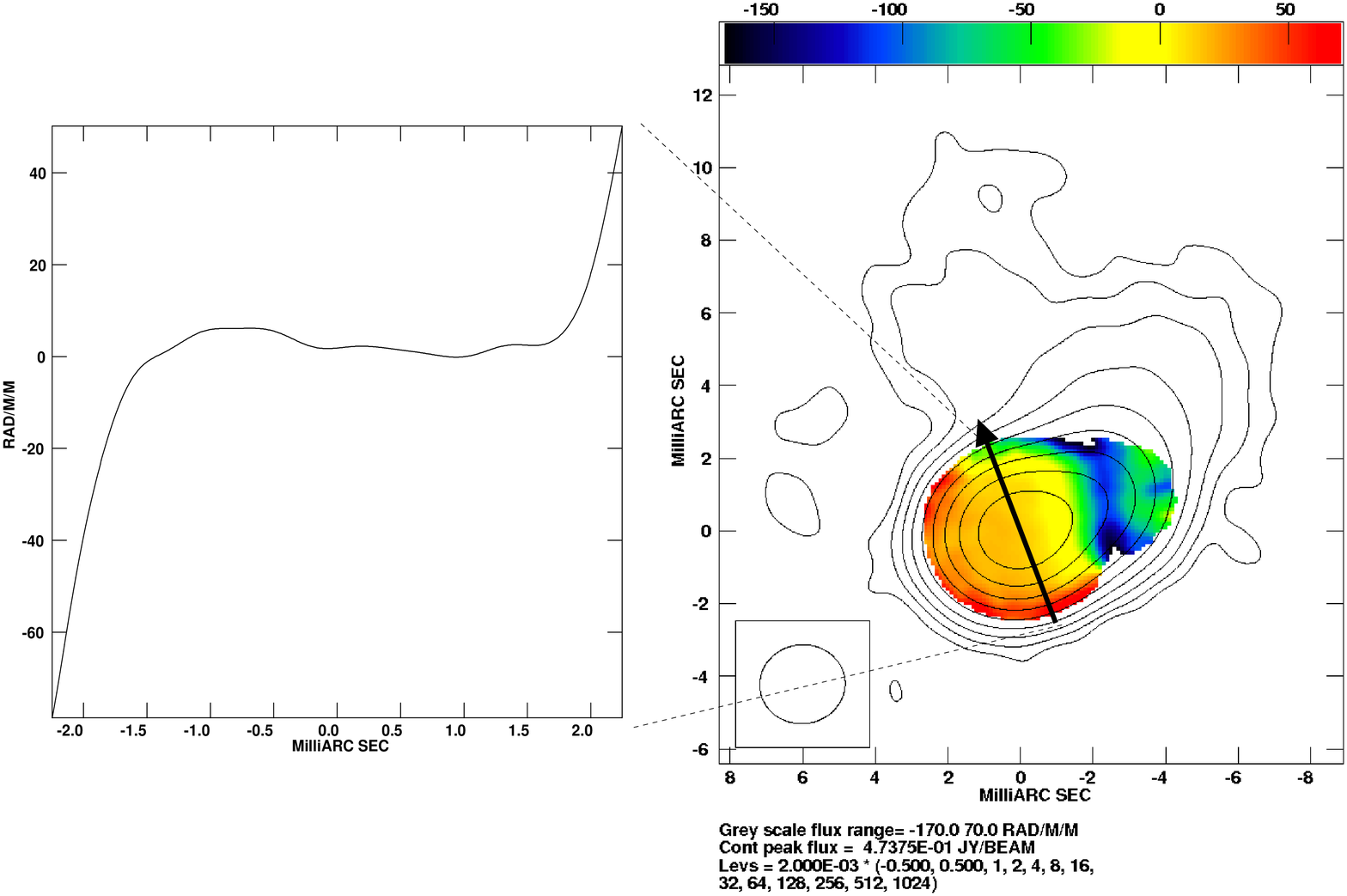}
\caption{RM map of 1749+701 showing a transverse RM
 gradient in the core/inner jet. The accompanying panels shows a slice of the 
 RM distribution across the core. Although it is not visible in our RM image,
 an opposing RM gradient in the jet is revealed in the 18cm RM map of
 Hallahan \& Gabuzda (these proceedings).}
 \end{center}
 \end{minipage}
\hspace{0.5cm}
 \begin{minipage}[t]{0.47\linewidth}
 \begin{center}
 \includegraphics[width=1\linewidth,clip]{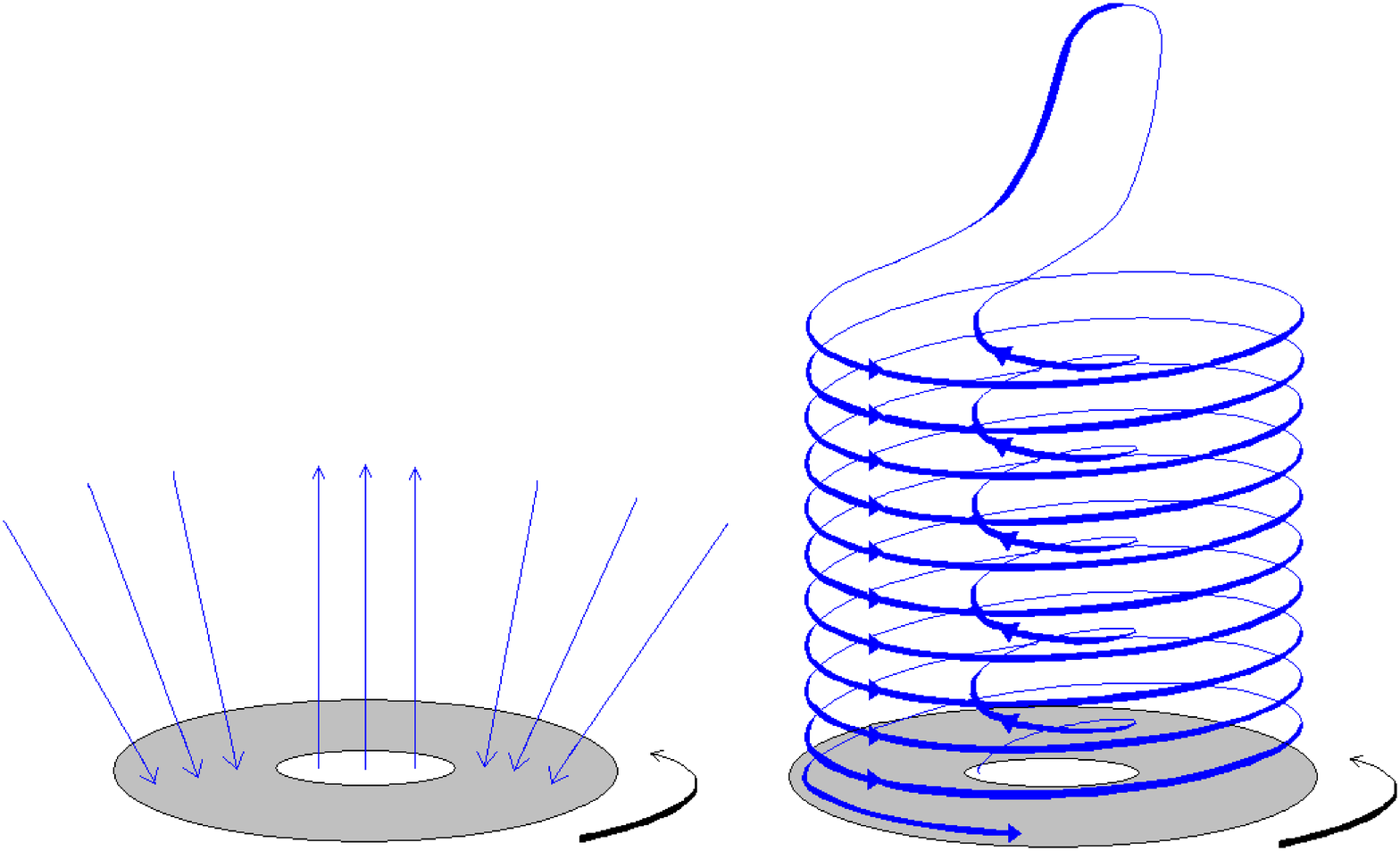}
 \caption{Illustration showing how the magnetic field lines from a magnetic tower model can get wound up as an ``inner'' and ``outer'' helix, as a result of the differential disk rotation.The first figure shows the direction of the magnetic field lines in this model, whereas the second follows the path of one of these magnetic field lines that gets wound up in a ``nested helical field'' with the ``inner'' helix less tightly wound than the ``outer'' helix.}
 \end{center}
 \end{minipage}
 \end{figure}

\section{Reversal of RM gradients in the core and jet}
The cm-wavelength cores of AGN that we observe with VLBI are located at some 
distance from the central black hole \cite{Mar02}. The location of the 
optically thick ``core'' ($\tau$=1 surface) along the jet outflow is wavelength 
dependant; the observed VLBI core is usually only partially optically thick, 
since the resolution provided is insufficient to fully isolate the optically 
thick core (base of the jet). The observed core thus includes emission from 
the inner jet of the source, and transverse RM gradients observed in the core 
region essentially correspond to the RM distribution in the innermost jet.

We have previously discussed several AGN that show transverse RM gradients 
across their jets \cite{Gab08,Mah07}. Here, we consider 5 sources (0138-097,
0256+075, 0716+714, 1334-127 and 2155-155) for which we observe transverse 
RM gradients across both their cores and their jets, with the gradients in 
the cores directed opposite to those in the jets (Fig. 1). We also include 
1749+701 (Fig. 2), for which we have detected an RM gradient in the core region
only, while both this gradient and an oppositely directed gradient in the jet
are visible in the 18cm RM map of Hallahan \& Gabuzda (these proceedings). 

If these transverse RM gradients are associated with helical jet {\bf B}
fields, these results seem to require a change in the direction of
the azimuthal {\bf B} field with distance from the core. It is not obvious 
how this could come about, since the direction of the azimuthal {\bf B}
field is essentially determined by the direction of the poloidal {\bf B} field
and the direction of rotation of the central black hole and accretion disc,
which we would expect to remain constant. However, let us consider a slightly
more complex ``magnetic tower'' configuration, with poloidal flux and poloidal 
current concentrated around the central axis \cite{Lyn96,Nak06}. Meridional 
{\bf B}-field loops anchored in the inner 
and outer parts of the accretion disc become twisted due to the differential 
rotation of the disc, essentially giving rise to an ``inner'' helical 
{\bf B} field near the jet axis and an ``outer'' helical field further from 
this axis (Fig. 3). These two regions of helical field will be 
associated with oppositely detected RM gradients, and the total observed RM 
gradient will be determined by which region of helical field dominates the 
observed RMs along a given line of sight.  Depending on factors such as the
pitch angles of the helical fields, the decrease in the electron density and 
field strength with distance from the axis and the jet base and the viewing 
angle, a particular observed RM gradient could correspond to the inner or 
outer helix. Thus, a change in the direction of the observed 
transverse RM gradient between the core/innermost jet and more distant jet 
regions could represent a transition from dominance of the inner and outer 
helical B fields in the total observed RM. It seems to make intuitive sense
that we are seeing RM gradients due to the inner helix in the innermost
jet/core region and due to the outer helix further from the core, but numerical 
studies are required to verify this. 

Typically, we might expect the direction of the RM gradients in the core and 
jet (i.e., the regions whose net RM gradients are determined by the 
inner/outer helical fields) to remain constant in time. However, this type 
of nested helical-field structure could also give rise to changes in the direction of the observed RM gradients with time. In fact, we have also observed a source (1803+784) with RM gradients in both the core and
jet, with the gradient in the jet ``flipping'' over time \cite{Mah08}. 


Another possible interpretation of the observed oppositely directed core and 
jet transverse RM gradients could be that the direction of the azimuthal 
{\bf B} field changed as a result of torsional oscillations of jet \cite{Bis07}. 
Such torsional oscillations, which may help stabilize the jets, could thus 
cause ``flips'' of the azimuthal {\bf B} field with time, or equivalently 
with distance from the core, given the jet outflow. In this scenario, we expect 
that the direction of the observed transverse RM gradients may reverse from 
time to time when the direction of the torsional oscillation reverses; this
can be tested by multi-wavelength polarization monitoring of AGN displaying
core and jet RM gradients.

\section{Conclusion}

These results provide new evidence for the presence of helical {\bf B} fields 
wrapped around blazar jets, which give rise to the observed transverse RM 
gradients. While it makes sense that these RM gradients should evolve over 
time, it is not clear how they can ``flip`' with distance from the core, 
as we have observed for the 5 sources considered here. Do the jets have two nested helical magnetic fields in a magnetic tower structure, with the
inner and outer helices dominating the observed RM gradient at different
distances from the base of the jet? If so, how common is this phenomena? Or
are these jets undergoing torsional oscillations that change the direction
of the azimuthal {\bf B}-field component? If so, are these oscillations 
periodic or irregular? Further observational studies of AGN jets displaying
such RM-gradient reversals can potentially provide crucial information about
the detailed geometry of the magnetic fields in AGN jets and how they evolve,
as well as information about the jet dynamics and collimation.

We have just obtained new multi-wavelength polarization observations of a number of the
AGN considered here with these goals in mind.

This publication has emanated from research supported by the Research Frontiers 
Programme of Science Foundation Ireland. The National Radio Astronomy 
Observatory is operated by Associated Universities Inc.

\end{document}